\documentclass[%
 twocolumn, nofootinbib, superscriptaddress, 
 amsmath,amssymb,
aps,
prl,
nobalancelastpage
]{revtex4-2}

\usepackage{graphicx}
\usepackage{dcolumn}
\usepackage{bm}

\usepackage{lipsum}
\usepackage{amsmath}
\DeclareUnicodeCharacter{2212}{-}
\usepackage{amsthm}
\usepackage{float}
\usepackage{verbatim}
\usepackage{verbatim}
\usepackage{amssymb}
\usepackage{acronym}
\usepackage{algpseudocode}
\usepackage{algorithm}

\usepackage{tikz}
\usepackage{array}
\newcolumntype{P}[1]{>{\centering\arraybackslash}p{#1}}
\usepackage{rotating}
\usepackage{booktabs, multirow} 
\usepackage{soul}
\usepackage{caption}
\usepackage{subcaption}
\newcommand{\setlabel}[1]{\edef\@currentlabel{#1}\label}
\usepackage{cancel}
\usepackage[normalem]{ulem}
\usepackage{soul}
\usepackage{xcolor}

\begin{document}

\preprint{APS/123-QED}

\title{Nutation-Based Longitudinal Sensing Protocols for High-Field NMR With Nitrogen Vacancy Centers in Diamond}

\author{Declan Daly}
\affiliation{Department of Physics, University of Maryland, College Park, MD, USA}
\author{Stephen J. DeVience}
\affiliation{Quantum Technology Center, University of Maryland, College Park, MD, USA}
\author{Emma Huckestein}
\affiliation{Department of Physics, University of Maryland, College Park, MD, USA}
\author{John W. Blanchard}
\affiliation{Quantum Technology Center, University of Maryland, College Park, MD, USA}
\author{Johannes Cremer}
\affiliation{Quantum Technology Center, University of Maryland, College Park, MD, USA}
\author{Ronald L. Walsworth}
\affiliation{Department of Physics, University of Maryland, College Park, MD, USA}
\affiliation{Quantum Technology Center, University of Maryland, College Park, MD, USA}
\affiliation{Department of Electrical and Computer Engineering, University of Maryland, College Park, MD, USA}

\date{\today}

\begin{abstract}
Nitrogen vacancy (NV) centers in diamond enable nuclear magnetic resonance (NMR) spectroscopy of samples at the nano- and micron scales. However, at typical tesla-scale NMR magnetic field strengths, NV-NMR protocols become difficult to implement due to the challenge of driving fast NV pulse sequences sensitive to nuclear Larmor frequencies above a few megahertz. We perform simulations and theoretical analysis of the experimental viability of NV-NMR at tesla-scale magnetic fields using a new measurement protocol called DRACAERIS (Double Rewound ACquisition Amplitude Encoded Radio Induced Signal).  DRACAERIS detects the NMR sample's longitudinal magnetization at a much lower driven Rabi frequency, more suitable technically for NV detection. We discuss how pulse errors, finite pulse lengths, and nuclear spin-spin couplings affect the resulting NMR spectra.  We find that DRACAERIS is less susceptible to pulse imperfections and off-resonance effects than previous protocols for longitudinal magnetization detection.  We also identify reasonable parameters for experimental implementation.
\end{abstract}

\maketitle
\section{Introduction}

\begin{figure*}
\includegraphics[width=\textwidth]{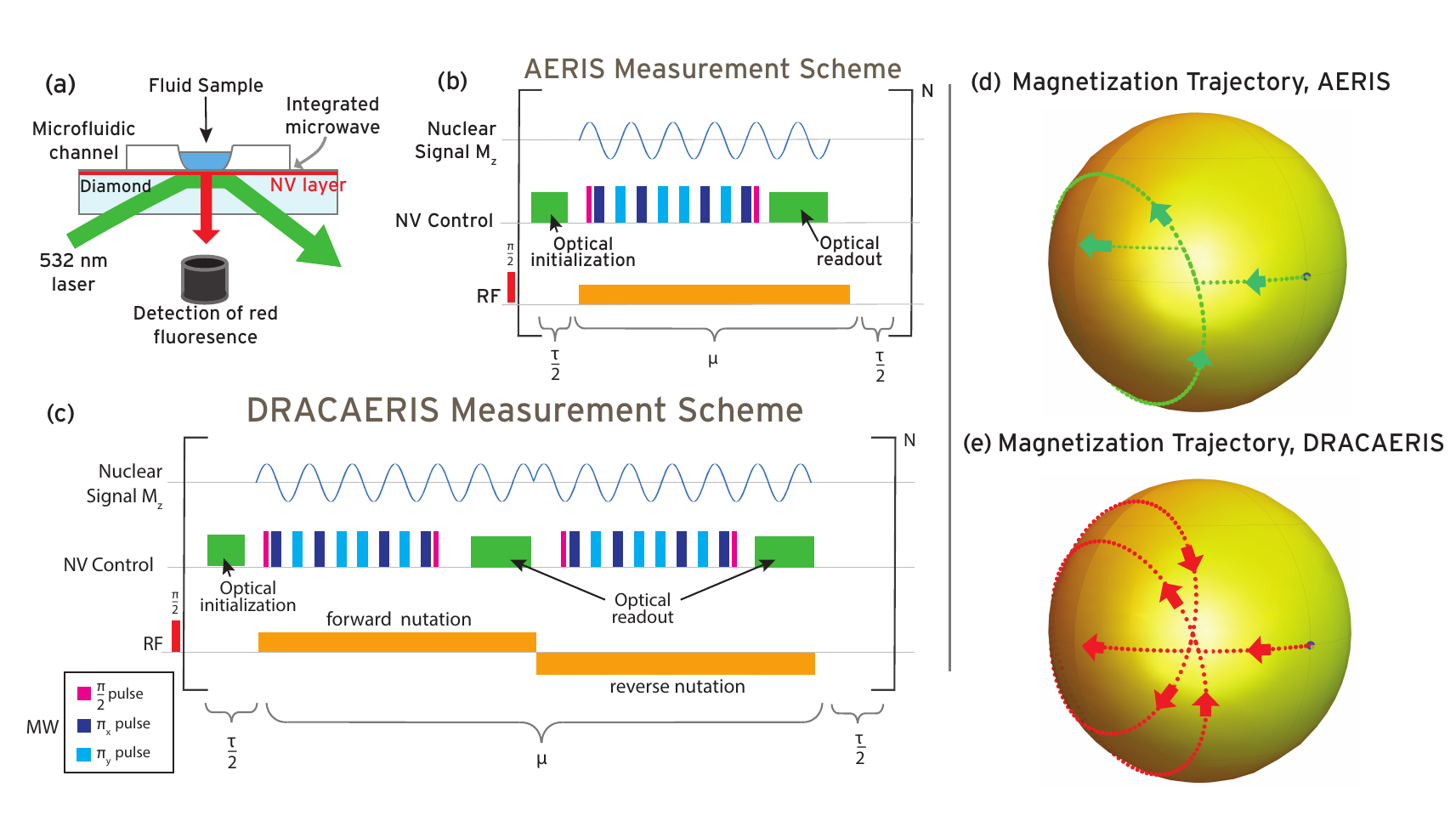}
\captionsetup{singlelinecheck=false,justification=raggedright}
\caption{\label{fig:wide}(a) Schematic of an ensemble NV-NMR apparatus.  A green laser optically excites an NV layer near the diamond surface, is totally reflected, and a camera or photodector collects the emitted NV red fluorescence. A fluid sample is in near contact with the NV layer, e.g., within a microfluidic apparatus.  An integrated coplanar waveguide applies microwaves to the NV ensemble. A nearby radiofrequency (RF) coil (not shown) is used to manipulate nuclear spins in the sample.  A bias magnetic field (not shown) provides Zeeman shifts to both the NV electronic spins and the sample nuclear spins.  (b) AERIS protocol from ref. \cite{munuera-javaloy_high-resolution_2023}.  Nuclear spins are initialized with a $\pi/2$ RF pulse and their precession is then measured via a series of $N$ acquisition steps. During each acquisition, the NVs are first initialized with an optical pulse. A $2\pi N$ RF nutation pulse with duration $\mu$ is applied to the nuclei. During this RF nutation pulse, a dynamical decoupling sequence such as XY8 is applied to the NV electronic spins via microwave (MW) pulses, followed by readout of the NV fluorescence. The time outside the RF pulse adds up to a delay $\tau$, during which the nuclear spins freely precess. (c) DRACAERIS protocol. Each acquisition now consists of two steps, a $2\pi N$ RF pulse on the nuclei driving forward nutation, followed by a second $2\pi N$ RF pulse with the opposite phase to rewind (i.e., reverse) the nutation. NV dynamical decoupling measurements are performed during both nutation pulses. Subtracting the second NV fluorescence signal from the first helps minimize common mode laser noise. Duration $\mu$ is defined as the total time for both $2\pi N$ RF nutation pulses. The box to the lower left gives the color-coded definitions of the different types of MW pulses applied to the NVs for both the AERIS and DRACAERIS protocols. Note that for both AERIS and DRACAERIS, typical operational conditions will have $\tau >\mu$, as discussed in the main text. (d) and (e) Bloch sphere diagrams highlighting errors caused by off-resonant driving of the nuclear spin magnetization.  In (d), the green trajectory shows the nuclear magnetization vectors before, during, and after the AERIS nutation pulse.  When the pulse is completed, the longitudinal magnetization vector $M_z$ does not return to the \textit{x-y} plane of the Bloch sphere.  In (e), the red trajectory shows that the DRACAERIS protocol corrects to leading order for these off resonance driving effects, and resets $M_z$ to zero after the two RF nutation pulses.}
\end{figure*}

Nitrogen-vacancy (NV) centers in diamond are widely used for micro- and nanoscale magnetic sensing applications due to their long electronic spin coherence times, optical spin-state preparation and readout at room temperature, and straightforward technical implementation \cite{zhang_diamond_2020, DegenReview,LevineReview}.  With dynamical decoupling sequences, NVs can be made sensitive to oscillating (AC) magnetic fields from nearby spins, e.g.,\,nuclear magnetic resonance (NMR) signals of small numbers of nuclei near the diamond surface \cite{Mamin2013,Staudacher2013,DeVience2015,Loretz2014,Rugar2015}. Using coherent sensing protocols, NV-NMR can achieve sensitivity better than $30 \text{ pT}/\sqrt{\text{Hz}}$ (for NV ensembles) and spectral resolution {\raise.17ex\hbox{$\scriptstyle\sim$}}1 Hz \cite{glenn_high-resolution_2018,aslam_nanoscale_2017,qdyne}, enabling high-resolution NMR spectroscopy of microscale sample volumes  \cite{glenn_high-resolution_2018,Kehayias2017,Bucher_2020}. Camera-based detection of ensemble NV fluorescence can also enable wide-field NMR spectral imaging with high spatial resolution \cite{DeVience2015}, providing a new tool for chemical analysis in the biological and physical sciences. A simplified schematic of an ensemble NV-NMR apparatus is shown in Fig.\ \ref{fig:wide}a.

Distinguishing chemical species via conventional NMR spectroscopy requires chemical shift resolution that is only achievable 
at high applied (bias) magnetic fields (typically $\sim$1 T or larger). To detect a sample NMR signal (produced by the sample's transverse nuclear magnetization oscillating at its Larmor frequency), 
a dynamical decoupling sequence is applied to the NVs with a pulse spacing on the timescale of half the nuclear Larmor precession period. At tesla-scale bias-field strengths, the NMR precession period is on the order of tens of nanoseconds, requiring that strong microwave (MW) pulses be applied to the NV electronic spins at high carrier frequencies \cite{aslam_nanoscale_2017}. For example, for a 1 T bias field, NV-NMR detection of a proton NMR signal requires that MW pulses be applied to the NVs at a carrier frequency $\sim$25 GHz, with a repetition rate of $\sim$85 MHz (given by the proton Larmor frequency) and a Rabi frequency $\sim$500 MHz (to approximate instantaneous pulses in the NV sensing protocol). 
Applying such large amplitude, high frequency MW pulses while also maintaining sufficient field homogeneity over the diamond surface is technically daunting and has yet to be realized for an ensemble NV-NMR system.

To address this challenge, an alternative measurement protocol was proposed 
called AERIS (Amplitude-Encoded Radio Intensity Signal) \cite{munuera-javaloy_high-resolution_2023}. Rather than detecting transverse NMR magnetization oscillating at the Larmor frequency, AERIS detects the sample's longitudinal nuclear magnetization oscillating at a much lower drive frequency, i.e.,
the nuclear Rabi frequency, which can be tuned to a technically optimal frequency for NV detection protocols (e.g., dynamic decoupling), typically hundreds of kHz to a few MHz. The AERIS protocol, shown in Fig.\ \ref{fig:wide}b, begins with an initial $\pi/2$ pulse, followed by a series of $2\pi N$ nutation pulses separated by regular intervals $\tau$, all applied to the sample nuclear spins. During these nutation pulses, the nuclear spins are driven at the chosen Rabi frequency while dynamical decoupling magnetometry is simultaneously performed via the NVs, yielding a series of NMR amplitude measurements over time. For each time point, the amplitude of the longitudinal magnetization signal depends on the relative phase between the nuclear spins and the phase of the nutation pulse, a phase shift that builds up during the inter-pulse intervals due to the frequency difference between the nuclear spin dynamics and the reference oscillator. The resulting longitudinal magnetization signal is expected to contain information about the nuclear spin sample analogous to that of a conventional transverse magnetization NMR signal detected inductively with a heterodyne circuit. 
Likewise, the Fourier transform of the resulting timecourse using the AERIS protocol is expected to be comparable to a conventional NMR spectrum, albeit with the ability provided by NVs to detect very small samples. 

As outlined above, AERIS provides an elegant solution to some challenges of high-field NV-NMR. However, non-ideal experimental conditions and sample properties may degrade its performance. For example, the finite bandwidth of pulses and the possibility of imperfect pulse rotations can lead to undesired dynamics of the nuclear spins. Finite lengths of the induced rotations also affect spin-spin dynamics within real sample molecules where J-couplings are present, thereby modifying the NMR spectra. In this work, we present an adapted protocol (DRACAERIS) designed to alleviate these effects and discuss how to interpret the resulting spectral data in relation to conventional NMR spectroscopy.




\section{DRACAERIS Protocol}
\setlabel{DRACAERIS Protocol}{sec:DracProtocol}

The AERIS protocol \cite{munuera-javaloy_high-resolution_2023} involves downsampling nuclear magnetization signals via longitudinal detection of the sample spins during repeated $2\pi N$ RF nutation pulses.  However, this proposal includes simplifications that are generally not applicable to real NMR samples, most importantly neglecting chemical shift differences and J-couplings during the nuclear drive pulses. Also, in practical applications pulses are not perfect throughout the sample because of imperfect calibration and field inhomogeneities. Since the AERIS protocol relies on repeated application of these pulses, even small pulse errors can have a large cumulative effect. For example, these errors break the underlying assumption that during the free precession period $\tau$ the nuclear spins are in the X-Y plane, which can lead to incomprehensible NMR spectra.


   

\begin{figure}
\centering
\includegraphics[width = 0.9\columnwidth]{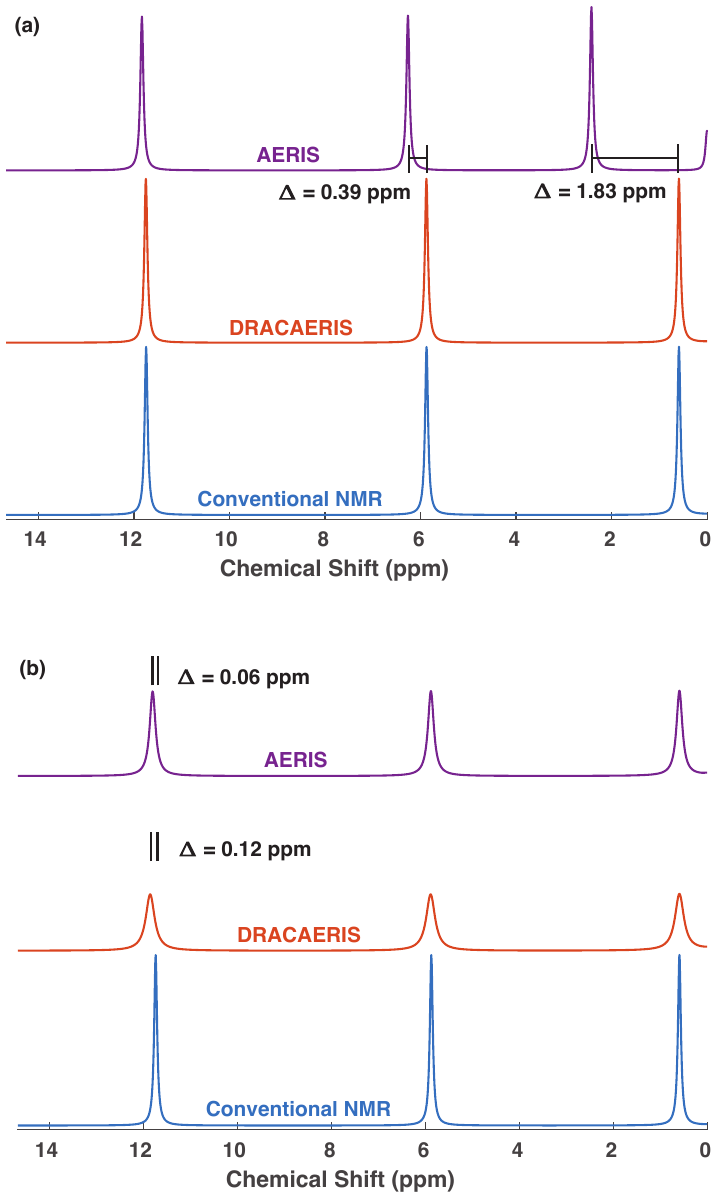}
\captionsetup{justification=raggedright, singlelinecheck=false}
\caption{\label{magnetization error}Longitudinal magnetization detection by an NV sensor produces spectral line shifts, in comparison to conventional NMR detection of transverese magnetization, in the presence of pulse errors and off-resonant spins. Here conventional, AERIS, and DRACAERIS detection of an NMR sample is simulated.  The sample contains a system with three types of uncoupled spins in a 1 T bias field with chemical shifts of 25, 250, and 500\,Hz ($\mathbf{\delta} = 0.59, 5.9, 11.7$\,ppm.), and a coherence time T$_2$ = 1 s for each spin type.  (a) With a 2\% pulse error, AERIS produces a significant shift $\Delta$ in some spectral lines, while DRACAERIS does not. Here $\tau = 800 \,\mu$s, $\mu = 20 \,\mu$s for AERIS and $\mu = 40 \,\mu$s for DRACAERIS, with Rabi frequency $\Omega = 200$\,kHz. (b) Even with perfect pulses, small errors can arise due to off-resonance effects if the Rabi frequency $\Omega$ is too low. Here $\tau = 800 \,\mu$s, $\mu = 800 \,\mu$s for AERIS and $\mu = 1600 \,\mu$s for DRACAERIS, with $\Omega = 5$\,kHz. The different linewidths reflect the significantly longer total measurement time when $\mu$ is large (see also Fig.\ 4).}
\end{figure}



In order to address pulse errors, the DRACERIS protocol includes a key modification: every $2\pi N$ ``forward'' RF nutation pulse is immediately followed by an equivalent ``rewinding'' (or reverse) pulse of opposite phase, illustrated in 
Figure\,1c. The rewinding pulse compensates to leading order for the effects of pulse errors and returns the spins to the X-Y plane.  We call this protocol DRACAERIS (Double Rewound ACquisition Amplitude Encoded Radio Induced Signal).  
To illustrate schematically the effect of pulse errors, Figure\,1d shows the magnetization of an off-resonant nuclear spin during one cycle of the AERIS sequence. Following the nutation pulse, magnetization is shifted out of the transverse plane. In contrast, using DRACAERIS (Figure\,1e), magnetization is shifted out of the transverse plane following the first (forward) nutation pulse, but is returned there by the second (rewinding) pulse. 

Pulse errors can cause significant spectral shifts that make NMR spectral identification difficult. Figure\,2a shows simulated measurements of three uncoupled spins with differing chemical shifts using conventional NMR detection, as well as NV-NMR with AERIS and DRACAERIS in a 1\,T field and with a 2\% error in the nutation rate used for AERIS and DRACAERIS. The extra phase buildup due to pulse errors causes AERIS measurements to shift toward higher frequencies, with the largest effects occurring for the lowest frequency NMR spectral lines. For example, the spin at a 25\,Hz chemical shift ($\delta =$ 0.59\,ppm) would be measured at 103\,Hz ($\delta =$ 2.42\,ppm) using AERIS,  corresponding to a spectral line error $\Delta = 1.83$ ppm. In comparison, DRACAERIS eliminates this effect and produces the correct NMR spectrum.

For weak nutation pulses, AERIS also produces small spectral shifts in off-resonance nuclear spins, an effect caused by limited 
nutation pulse bandwidth. Figure\,2b shows simulated measurements of the same spins as in Figure 2a, but for perfect pulses and a weak nutation pulse with a Rabi frequency of just 5\,kHz. In this example of an extreme case of weak pulses and thus small bandwidth, the AERIS nutation time $\mu$ equals the free precession time $\tau$, and the nutation Rabi frequency is only about ten times larger than the chemical shift of the farthest NMR spectral line. As a result, off-resonant spectral lines are shifted slightly, and the effect is larger for spins that are further off-resonance. The maximum shift using these parameters is about 0.5\%.

DRACAERIS does not correct for these effects, and in fact produces about twice as large a shift due to the extra nutation pulse. Fortunately, off-resonance effects are much smaller for typical experimental parameters. To achieve high sensitivity with either protocol, nuclear spin driving must be strong enough to produce multiple Rabi oscillations within the NV dephasing time $T_2$; also, shorter MW pulse spacing during the dynamical decoupling sequence improves NV $T_2$. Given a typical NV ensemble $T_2$ of 20\,$\mu$s for an XY8 sequence \cite{Droid}, a minimum nuclear spin RF nutation pulse Rabi frequency $\Omega=$ 200\,kHz is necessary to create four periods of nuclear oscillations. Adding in rewinding for DRACAERIS gives a total nutation pulse time $\mu$ of at least 40\,$\mu$s.  An additional constraint arises from the required sampling bandwidth. For example, to cover roughly 15\,ppm of a proton NMR spectrum at 1\,T, one would need a sampling frequency of at least $f_{sr}=1.25$\,kHz to achieve the Nyquist limit. Thus $\tau = 1/f_{sr}$ must be shorter than $\sim$800\,$\mu$s. A typical ratio $\tau/\mu$, i.e., the amount of free precession versus acquisition time, is therefore $<$ 20. In this scenario the nutation Rabi frequency is about 300 times stronger than the largest chemical shifts, and off-resonance effects for AERIS or DRACAERIS are much smaller than the NMR linewidth for a typical sample.


Because data acquisition is only possible during $\mu$, sensitivity is expected to scale by a factor of $\sqrt{\mu/(\mu+\tau)}$ compared with NV-NMR methods using near continuous acquisition, such as CASR \cite{glenn_high-resolution_2018}. However, utilizing DRACAERIS together with a quantum logic enhanced readout scheme, as described in \cite{Droid}, would allow NV readout to continue during the nuclear spin free evolution period $\tau$, offering sensitivity closer to continuous acquisition without changing the nutation duration $\mu$.

\subsection{MAGNETIC FIELD RESCALING} 
\setlabel{MAGNETIC FIELD RESCALING}{sec:axis_scaling}

A third cause of measurement error for NV-NMR longitudinal sensing protocols is nuclear spin-spin coupling in the target sample. During conventional inductive NMR detection, the signal resulting from transverse magnetization  $\langle M_y\rangle$ is sampled at a repeated time interval $\tau$. For conventional inductive detection, $\tau$ tracks the free precession time of the nuclear spins. However, for AERIS and DRACAERIS, the signal is measured at intervals of $\tau + \mu$. If evolution of the sample nuclear spin system were suspended during the acquisition time $\mu$, then the NMR spectrum created by the three techniques would be identical when using $\tau$ to measure free precession time. This condition applies if the sample nuclei are described only by the chemical shift Hamiltonian, since during strong driving (i.e.,\ nutation pulse Rabi frequency $\Omega$ much larger than chemical shift differences $\delta \omega$) chemical shift evolution is suppressed. In this case the spin system at the end of the nutation pulse is nearly identical to its state at the beginning of the pulse. 

However, for coupled spin systems, J-coupling terms of the Hamiltonian are not suppressed by strong nutation driving; i.e., the J-coupling Hamiltonian is effective for the entire time $\tau + \mu$. In this case, if one uses $\tau$ as the measure of evolution time, our simulations indicate that the chemical shift frequencies match the conventional spectrum, but the splittings caused by J-coupling are too large (Figure\,3a). If the sampling time step is instead considered to be $\tau + \mu$, then our simulations show that the J-coupling splittings have the correct value, but the chemical shifts are too small (Figure\,3b). 

These systematic errors can be corrected with a straightforward modification to the analysis of the longitudinal sensing data. From simulations and average Hamiltonian theory, we find that for all non-zero $\mu$, the AERIS or DRACAERIS spectrum matches the conventional NMR spectrum acquired at a bias magnetic field rescaled by $\tau / (\tau+\mu)$ (where $\tau$ is again the measure of free precession time for the time-domain signal). For example, if $\tau/\mu = 1$, a DRACAERIS spectrum acquired at 1 T would match the conventional spectrum acquired at 0.5 T (Figure\ 3c).

\begin{figure}[h!]
\includegraphics[width = 0.9\columnwidth]{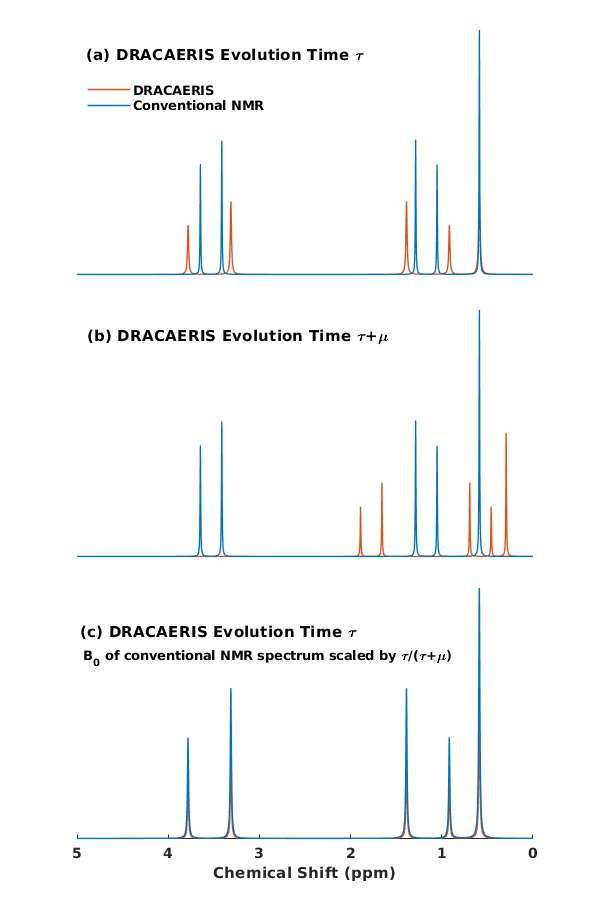}
\captionsetup{singlelinecheck=false,justification=raggedright}
\caption{Longitudinal magnetization detection by an NV sensor using DRACAERIS produces spectral distortions for coupled spins, which can be corrected using magnetic field rescaling. Here, conventional and DRACAERIS detection of an NMR sample is simulated.  The sample contains three types of nuclear spins at a 1 T bias field. Spin 1 is uncoupled with chemical shift $\mathbf{\nu} = 25$\,Hz ($\mathbf{\delta} = 0.59$\,ppm). Spins 2 and 3 have $\mathbf{\nu} = 50, 150$\,Hz ($\mathbf{\delta} = 1.17, 3.52$\,ppm) and have $J = 10$\,Hz scalar coupling. (a) When the free precession time $\tau$ is used as the evolution time for the frequency axis, DRACAERIS yields spectral lines with the correct chemical shift but incorrect splitting due to J-coupling. (b) When the total evolution time $\tau+\mu$ is used, DRACAERIS gives the correct splitting due to J-coupling but incorrect chemical shifts. (c) A match is achieved for a conventional NMR spectrum at a magnetic field rescaled by $\tau/(\tau+\mu)$. Due to the longer experimental time for longitudinal sensing, the DRACAERIS spectral features have lower amplitudes and broader linewidths than conventional NMR. }
\end{figure}

This effect can be significant for the NV-NMR spectra of real molecules. Figure\,4 shows an example simulation for ethanol. Both low and high field spectra show better equivalence with conventional NMR spectra as $\tau/\mu$ increases, particularly when $\tau/\mu \geq 10$.  At 1\,T and $\tau/\mu =1$, the spectra begin to show extra splittings typical of those acquired at much lower field strengths, which result from higher-order effects of J-coupling. Figure\,5 shows that scaling the reference spectrum's field by $\tau/(\tau+\mu)$ produces a match with DRACAERIS for all spectral features.

\begin{figure*}
\label{why does this label break}
\includegraphics[width = \textwidth]{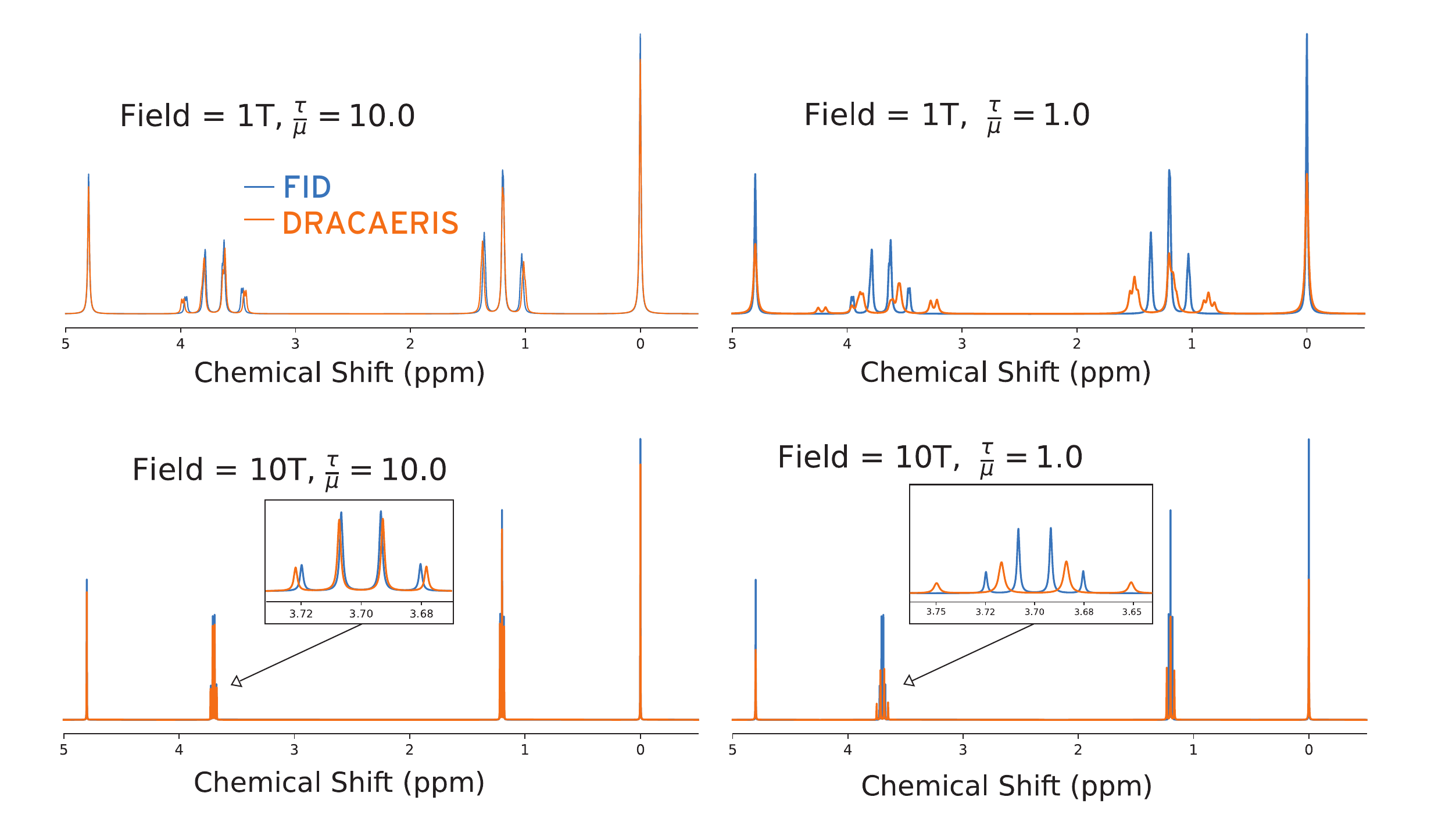}
\captionsetup{singlelinecheck=false,justification=raggedright}
\caption{\label{ethanol spectra} Simulated ethanol NMR spectra using conventional inductive and longitudinal DRACAERIS detection disagree due to the effects of scalar coupling between the sample spins. These effects become more pronounced for smaller ratios of $\tau/\mu$. The insets illustrate that the magnitude of disagreement are similar for 1\,T and 10\,T bias magnetic fields, although at 1\,T extra splittings also appear due to higher-order effects. Uncoupled spins are unaffected. The 0\,ppm line is a tetramethylsilane reference. Here $\tau = 1$\,ms, $\Omega = 20$\,kHz, and nuclear spin $T_2 = 1$\,s.}
\end{figure*}

\section{Discussion}


Longitudinal detection is an elegant way of shifting an NMR signal to a more convenient detection frequency and has been implemented in magnetic resonance in the past \cite{schweiger,whitfield}, including for some of the first NV-NMR measurements \cite{Mamin2013}. However, the required driving pulses are subject to errors in amplitude and frequency in any realistic NMR apparatus, due to both intrinsic factors such as off-resonant spins and extrinsic factors like $B_1$ inhomogeneity. A common way to correct for pulse errors is to perform a second operation that reverses the leading order effect of an error; this can occur by either cycling the phases of pulses over the course of a measurement or by directly rewinding the error during the measurement step. The original description of AERIS \cite{munuera-javaloy_high-resolution_2023} proposed an error-correction protocol, which like DRACAERIS performs a rewinding immediately before any free-precession can occur under imperfect conditions. However, the protocol in \cite{munuera-javaloy_high-resolution_2023} performs the forward and reverse pulses within a single modified NV magnetometry sequence, while DRACAERIS instead splits this correction between two separate NV magnetometry sequences. The DRACAERIS approach has two key benefits. First, it refrains from altering the NV sensing protocol (such as XY8), which could corrupt its error-correction properties for NV pulses. Second, it provides a differential measurement that reduces the effects of laser intensity noise \cite{bar-gill_solid-state_2013}. 

\begin{figure*}
\includegraphics[width = \textwidth]{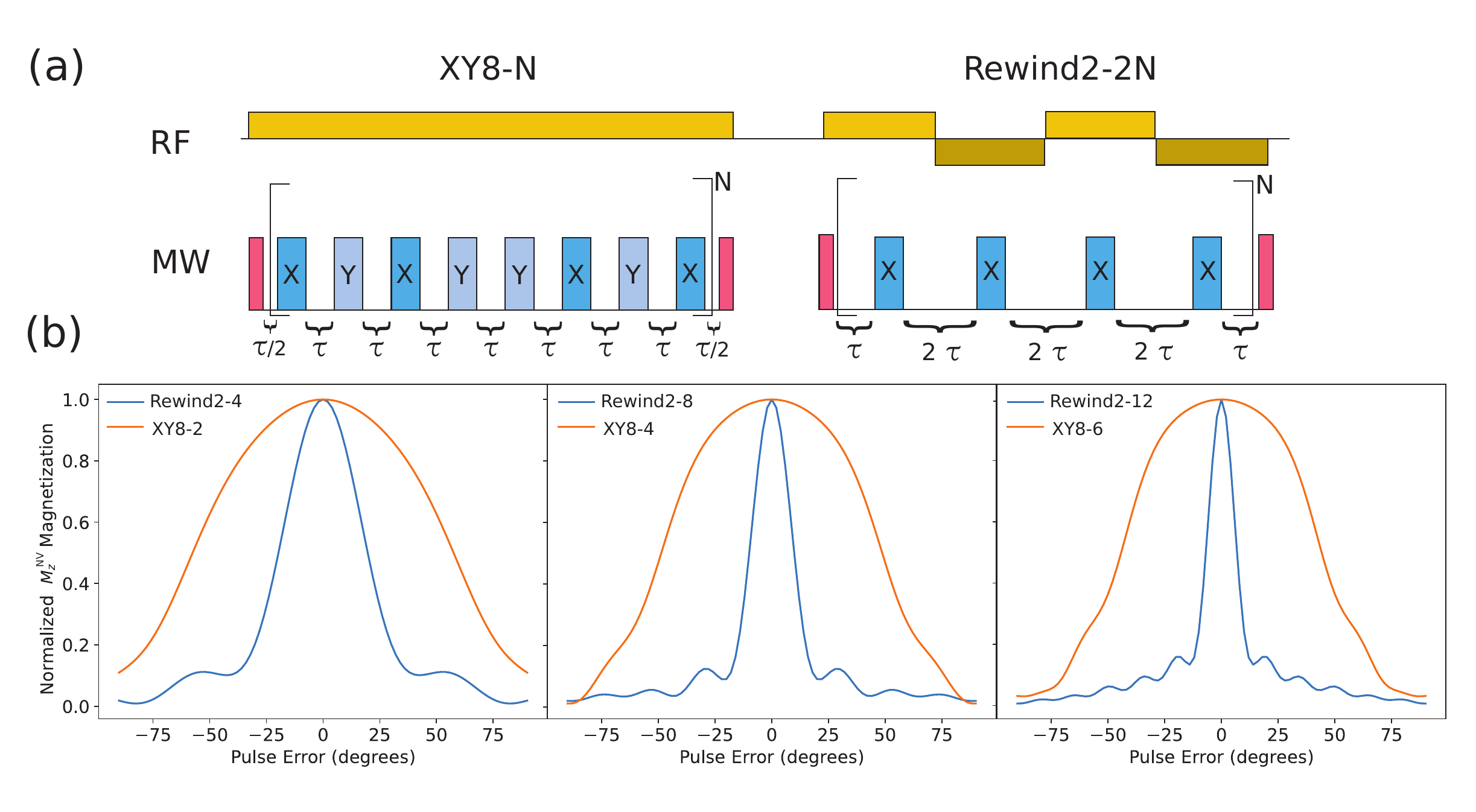}
\captionsetup{singlelinecheck=false,justification=raggedright}
\caption{\label{rewind comparison}Simulations indicate that DRACAERIS measurements of longitudinal signals are much less sensitive to MW pulse errors than the rewound acquisition version of AERIS \cite{munuera-javaloy_high-resolution_2023}. (a) Here, we compare the performance of multi-cycle XY8-N sequences used in DRACAERIS to the modified AERIS sequence over multiple cycles (Rewind2-2N). These two sequences have the same phase accumulation time. (b) Simulations of the normalized NV ensemble longitudinal magnetization (proportional to the measured NV fluorescence) show that XY8 sequences perform significantly better than Rewind2 in the presence of pulse errors, particularly for large numbers of repetitions. In the absence of pulse error, the sequences perform about equally (within 5\%, results shown here are normalized to the maximum measured signal to make bandwidth comparisons easier). Simulations are for a 50\,pT amplitude signal driven at 1 MHz and measured using an NV ensemble with Gaussian-averaged $T_2^*=100$\,ns, a typical value \cite{jennythesis}.}
\end{figure*}

To quantify the advantage of DRACAERIS in the presence of MW pulse errors, we simulate and compare the performance of the modified AERIS protocol \cite{munuera-javaloy_high-resolution_2023} and the forward nutation portion of a single DRACAERIS measurement. We use ideal NMR signals but vary the lengths of MW pulses applied to the NV ensemble to model common rotational pulse error.  We take several steps to ensure the simulation reflects meaningful measurements.  To improve contrast, it is beneficial to use many cycles of XY8, designated XY8-N, where N is the number of cycles. To create a fair comparison, we also extend the modified AERIS sequence over multiple cycles, which we designate as the Rewind2-2N protocol. A single Rewind2 and a single XY4 cycle have the same sensing time, similar sensitivity, and similar levels of degradation due to pulse errors. However, as shown in Fig. \ref{rewind comparison}, the XY8 protocol used in DRACAERIS maintains its sensitivity much better than Rewind2 as the number of cycles and sensing time increases. This superior performance is because extended Rewind2 sequences have vanishing error correction benefits of alternating X and Y phases, making them analogous to CPMG sequences in the long sequence duration limit.

Regardless of the measurement scheme utilized, a rewind nutation is critical to the performance of any longitudinal protocol.  Without rewind correction, a consistent pulse error applied to a resonant nuclear spin causes evolution along the X-Z plane, which leads to an oscillating X-axis magnetization that appears as a frequency shift in the NMR spectrum (see Supplementary Material \cite{DRACAERIS_SM}). 
For off-resonant spins, this effect is somewhat suppressed once a spin precesses toward 180$^\circ$ around the Bloch sphere. By keeping the nuclear spins sufficiently far from resonance, it might be possible to minimize the effects of pulse errors without resorting to rewinding. However, this approach wastes bandwidth and increases the effects of shifts caused by off-resonant excitation.

Even after correcting for pulse errors using DRACAERIS, it is important to account for spectral effects caused by the nutation pulses. NMR spectra typically have features produced by both chemical shifts ($\delta$), which are field-strength dependent, as well as field-independent scalar couplings ($J$) between spins. The interplay between these effects causes the spectra to change significantly as a function of magnetic field. At sufficiently high bias field strengths, such that chemical shift differences are significantly larger than scalar couplings ($\Delta \delta \gg J$), chemical shifts dominate the nuclear spin Hamiltonian and only the secular coupling term $2\pi J_{jk} I_{z, j} I_{z, k}$ is significant. In this case, NMR spectra typically consist of multiplets with splitting equal to $J$, separated by larger chemical shift differences. A typical example is the ethanol NMR spectrum at a 1 T bias field, as shown in Fig.\ 4. At lower bias fields where $\Delta \delta \le J$, higher-order effects caused by the non-secular coupling terms $2\pi J_{jk} (I_{x, j} I_{x, k}+I_{y, j} I_{y, k})$ also become significant. 


The effective chemical shift exhibited by the NMR sample also can be controlled by the measurement protocol.  For example, the chemical shift can be suppressed during sufficiently strong pulses or by spin locking applied to nuclear spins, as well as by time reversal through one or more spin echoes. Morris et al.\ demonstrated these effects in inductively-detected NMR with a chemical shift scaling protocol in which an XY dynamical decoupling sequence was performed between signal acquisitions \cite{morris_css}. The result is an NMR spectrum acquired at bias field $B_0$ matching a conventional NMR spectrum acquired at $\tau_2/(\tau_1+\tau_2)B_0$, where $\tau_1$ is the dynamical decoupling time and $\tau_2$ is the spin free precession time. This result matches our field scaling for DRACAERIS, consistent with the effects of dynamical decoupling and strong spin driving being analogous. Additionally, as in \cite{morris_css}, altering the nutation duration may provide access to different spectral behavior in a high-field setting.

\begin{figure}
\includegraphics[width = \columnwidth]{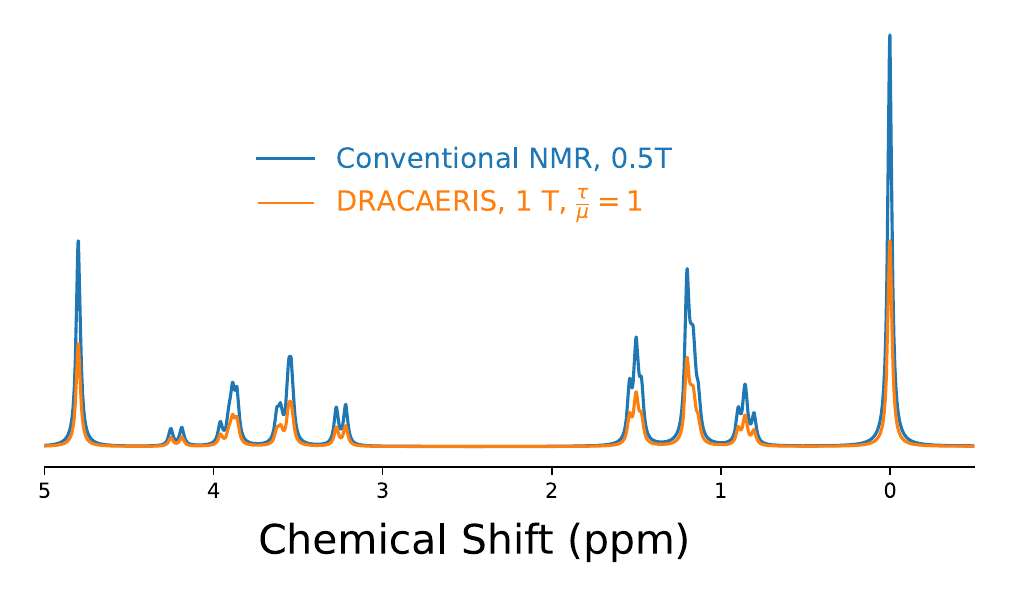}
\captionsetup{singlelinecheck=false,justification=raggedright}
\caption{\label{field comparisons} Simulated ethanol DRACAERIS NMR spectrum at 1\,T bias field with $\tau/\mu=1$ is reproduced from Figure 4 and compared against an inductive NMR spectrum simulated for 0.5\,T. There is good agreement among the locations of all spectral features. Nuclear spin decoherence causes the DRACAERIS spectral features to have lower amplitudes and somewhat broader linewidths due to the additional experiment time required for the longitudinal sensing protocol.}
\end{figure}


\section{Conclusion}




We perform simulations of longitudinal detection techniques for NV-NMR, and find that these techniques can create measurement errors due to pulse imperfections and off-resonance effects. We introduce a new longitidunal detection protocol called DRACAERIS (Double Rewound ACquisition Amplitude Encoded Radio Induced Signal) that simulations show can greatly suppress the errors from imperfect pulses. Additionally, we provide a model based on quantum simulations and confirmed by Average Hamiltonian Theory to help interpret longitudinally detected NMR spectra. Our results indicate that magnetic field rescaling can easily correct for predictable spectral distortions, due to the effects of J-couplings between sample nuclear spins, present in such spectra.  This study provides a basis for experimental realization of the DRACAERIS protocol, which we intend to pursue in future work.

\section{Acknowledgements}
We thank Niko Reed for assistance with figure creation.  This work is supported by, or in part by, the U.S. Army Research Laboratory MAQP Program under Contract No. W911NF1920181; the U.S. Army Research Office under Grant No. W911NF2120110;  the U.S. Air Force Office of Scientific Research under Grant No. FA9550-22-1-0312; the Gordon \& Betty Moore Foundation; and the University of Maryland Quantum Technology Center.

\section{Author Contribution}
D.D.\ initiated the investigation and then developed the DRACAERIS protocol together with E.H.\ and J.W.B.  D.D.\ and S.J.D.\, in consultation with J.W.B.\ and J.C.\, developed the theory and performed the simulations. R.L.W.\ supervised the work. All authors discussed the research and contributed to the final manuscript.  
\section{Methods}

To investigate how nuclear spins evolve under the AERIS and DRACAERIS protocols, a numerical simulation was created to track the density operator of the sample spins under the effects of different Hamiltonians during both the free evolution $\tau$ and nutation period $\mu$. In the rotating frame, these Hamiltonians can be defined as 
\begin{equation}
\hat{H}_{0} = \sum_{i}\omega_i \hat{I}_{z, i} + \sum_{j>k}2 \pi J_{jk} \bf{\hat{I}}_{j}\cdot \bf{\hat{I}}_{k}
\end{equation}
during free evolution and 
\begin{equation}
\hat{H}_{drive} = \hat{H}_{0} + \sum_{i} \Omega \hat{I}_{x, i}
\end{equation}
during nutation pulses, where $\omega_i$ represents the \textit{i}th chemical shift precession frequency in the rotating frame, $\bf{\hat{I}}$ and its components are the nuclear spin operators, $J_{jk}$ represents the spin-spin coupling constant between the \textit{j}th and \textit{k}th spins of the sample molecule, and $\Omega$ is the Rabi frequency of the nutation pulse. Since neither Hamiltonian is time dependent, one can use the simple time evolution operator $U(\hat{H}, t) = e^{i \hat{H} t}$ to evolve the density matrix of the sample spins under both free evolution $\tau$ and driving $\mu$ periods.  Due to chemical shifts, the resonance frequencies of different nuclei are not equal. This results in small deviations of the effective Rabi frequencies for each class of sample spin.  This is reflected in this formulation of the Hamiltonian, since the driving field $\Omega I_{x}$ is static in the rotating frame, and therefore is not resonant for any of the $\omega_i I_{z, i}$ terms unless $\omega_i = 0$.

At the beginning of each simulation, all nuclear spins begin along the $\hat{y}$ axis, representing the state following a $\pi/2$ excitation pulse along $-x$.  This state is represented by a density matrix $\rho$, and evolution under an arbitrary Hamiltonian $\hat{H}$ is calculated using the time-evolved state $\rho(t)=U(\hat{H}, t)^{-1}\rho \,U(\hat{H}, t)$. Excitation is followed by a series of repeated NV-NMR signal acquisition blocks. In each block, the free evolution Hamiltonian $\hat{H}_{0}$ is applied to the nuclei for duration $\tau/2$, the state is subjected to the nuclear driving Hamiltonian $\hat{H}_{drive}$ for nutation duration $\mu$, and finally the free evolution Hamiltonian $\hat{H}_{0}$ is applied to the nuclei for duration $\tau/2$ again. In an actual experiment, one or more AC magnetometry sequences are performed on the NVs during nuclear driving, and the resulting fluorescence produced by the NVs is directly related to the amplitude and phase of the longitudinal nuclear magnetization ($M^{NV}_z$).  For each acquisition block, we can therefore determine what the NV response would be by calculating $\langle M^{NV}_z \rangle = \text{tr}(\rho(t_{max}) \bf{\hat{I}_z})$ at the first maximum of the nutation oscillation $t_{max}$ (i.e.,\ a quarter of a period in a $2\pi$ RF pulse used for nuclear spin nutation). The signal acquisition blocks are repeated on the order of hundreds to thousands of times, matching the duration of the nuclear $T_2$. The Fourier transform of these acquisitions produces the longitudinally-detected NV-NMR spectrum.

The effects of nuclear spin decoherence are implemented by scaling the magnetization magnitude by a factor 
\begin{equation}
    \exp\left(-\frac{\tau+\mu}{T_2}\right) 
\end{equation}
after each acquisition step. In the case of conventional NMR, $\mu = 0$. Here we make the assumption that the nuclear spin decoherence times during the nutation pulse and free precession are similar, which is typically valid for liquid samples in a homogeneous field \cite{Slichter, Solomon}.

To perform the pulse error analysis shown in Figure \ref{rewind comparison}, we track the geometric position of the NV spin magnetization vector $\bar{M}$ as it accumulates phase in the presence of bias field $B_0$ and longitudinal nuclear spin signals produced by Rewind2 and DRACAERIS nutation pulses. Similar to our simulation of the nuclear spins, the magnetization vector of the NV ensemble is initialized in the transverse plane along $\hat{x}$ to reflect an initial MW $\pi/2$ pulse. The ensemble NV magnetization orientation is represented by the 2x2 density matrix of a two-level system.  MW $\pi$ pulses on the NVs are applied as simple geometric rotation operators around the $\hat{x}$ and $\hat{y}$ axes.  The degree of rotation induced by a $\pi$ pulse is varied from $\theta \, \epsilon \, [90, 270]$ degrees, corresponding to a set of pulse errors between -90 and +90 degrees.  At the end of the sequence, a final MW $\pi/2$ pulse projects the ensemble NV magnetization along $\hat{z}$ for measurement as the $\hat{z}$-projection  $\langle M^{NV}_z \rangle$, which is then plotted for each value of $\theta$.

To account for NV dephasing effects dictated by the characteristic time $T_2^{*}$, we repeated the above protocol for 100 different values of bias field intensity $B_0$ selected from a Gaussian distribution.  The width of the Gaussian distribution $\sigma = \frac{1}{\sqrt{2}\pi T_2^{*}}$ increases with shorter $T_2^{*}$ \cite{jennythesis}, and $\langle M^{NV}_z \rangle$ measurements for each value of $B_0$ are averaged together.



\bibliography{main.bib}

\appendix
\newcommand{\oneminus}{I_1^-}
\newcommand{\oneplus}{I_1^+}
\newcommand{\twominus}{I_2^-}
\newcommand{\twoplus}{I_2^+}
\newcommand{\onez}{I_{1, z}}
\newcommand{\twoz}{I_{2, z}}

\section{Appendix A: Average Hamiltonian Theory}

Using Average Hamiltonian Theory (AHT), we confirm the magnetic field scaling behavior observed in the simulated spectra. Consider a system in the rotating frame with two spin-1/2 nuclei interacting via spin-spin coupling $J_{12}$. The resonance frequencies are $\omega_1$ and $\omega_2$. The Hamiltonian is then
\begin{equation}
\hat{H}_0 = \omega_1 I_{z, 1} + \omega_2 I_{z, 2} + J_{12} \textbf{I}_1\cdot\textbf{I}_2.
\end{equation}
The driving Hamiltonian for this system is written in the following way:
\begin{equation}
\hat{H}_1 = \Omega(I_{x, 1} + I_{x, 2})
\end{equation}
The Hamiltonians during free precession and nutation pulses are then
\begin{equation}
\hat{H}_{int} = \begin{cases}
        \hat{H}_0 & \text{if } t \in [0, \tau]\\
        e^{-\Omega i (\textbf{I}_x)   (t-\tau )}\hat{H}_0
        e^{\Omega i (\textbf{I}_x) (t-\tau )} & \text{if } t \in [\tau, \tau + \mu]
    \end{cases}
\end{equation}
The average Hamiltonian for a single acquisition cycle is given by the integral
\begin{eqnarray}
\hat{H}_{avg} = \frac{1}{\tau + \mu} \left(\int_0^{\tau} H_{int} dt + \int_{\tau}^{\tau + \mu} H_{int} dt \right).
\end{eqnarray}
This integral evaluates to
\begin{eqnarray*}
    \hat{H}_{avg} = \frac{\tau}{\tau + \mu} 
                    \left( J_{12}\textbf{I}_1 \cdot \textbf{I}_2 + \omega_1 I_{1, z} + \omega_2 I_{2, z} \right) \\ +
                    \frac{\mu}{2(\tau + \mu)} 
                    \Biggl[J_{12}(\oneminus \cdot\twoplus + \oneplus \cdot \twominus + 2 (I_{1, z} I_{2, z})) \\ 
                    - \frac{2 i (\omega_1 (\oneminus - \oneplus) + \omega_2 (\twominus - \twoplus))\sin^2(\mu \Omega/2)}{\mu\Omega} \\
                    + \frac{2(\omega_1 \onez + \omega_2 \twoz)\sin(\mu \Omega)}{\mu\Omega}\Biggr]
\end{eqnarray*}
The terms divided by $\mu\Omega$ become zero if nutation pulses are perfect so that $\mu\Omega = 2\pi N$, i.e., at least one full nutation is performed. Even if pulses are imperfect, these terms are typically small, as they are suppressed by the factor $\Omega (\tau+\mu)$.  The remaining terms are:
\begin{eqnarray}
\hat{H}_{avg} = \frac{\tau}{\tau + \mu} 
                    \left( J_{12}\textbf{I}_1 \cdot \textbf{I}_2 + \omega_1 I_{1, z} + \omega_2 I_{2, z} \right) \\ +
                    \frac{J_{12} \mu}{2(\tau + \mu)} (\oneminus \cdot\twoplus + \oneminus + \cdot \twominus + 2 (I_{1, z} I_{2, z}))
\end{eqnarray}
Using $I_i^{\pm} = I_{i, x} \pm I_{i, y}$,

\begin{eqnarray}
\hat{H}_{avg} = \frac{\tau}{\tau + \mu} 
                    \left( J_{12}\textbf{I}_1 \cdot \textbf{I}_2 + \omega_1 I_{1, z} + \omega_2 I_{2, z} \right) \\ +
                    \frac{\mu}{\tau + \mu} (J_{12} \textbf{I}_1 \cdot \textbf{I}_2)
\end{eqnarray}
Adding terms together gives 
\begin{equation}
\hat{H}_{avg} = \frac{\tau}{\tau + \mu} (\omega_1 I_{1, z} + \omega_2 I_{2, z}) + J_{12}\textbf{I}_1\cdot \textbf{I}_2\quad.
\end{equation}
This first-order average Hamiltonian is for a spin system whose resonance frequencies have been scaled by factor $\tau/(\tau + \mu)$, a result that can also be obtained by scaling $B_0$ by the same factor. In the strong limit that $\mu \rightarrow 0$, this scaling factor disappears and the AHT first order result approaches the conventional NMR Hamiltonian.

\setcounter{figure}{0}

\renewcommand{\thefigure}{S\arabic{figure}}

\begin{figure*}[h!]
\includegraphics[width =.8\textwidth]{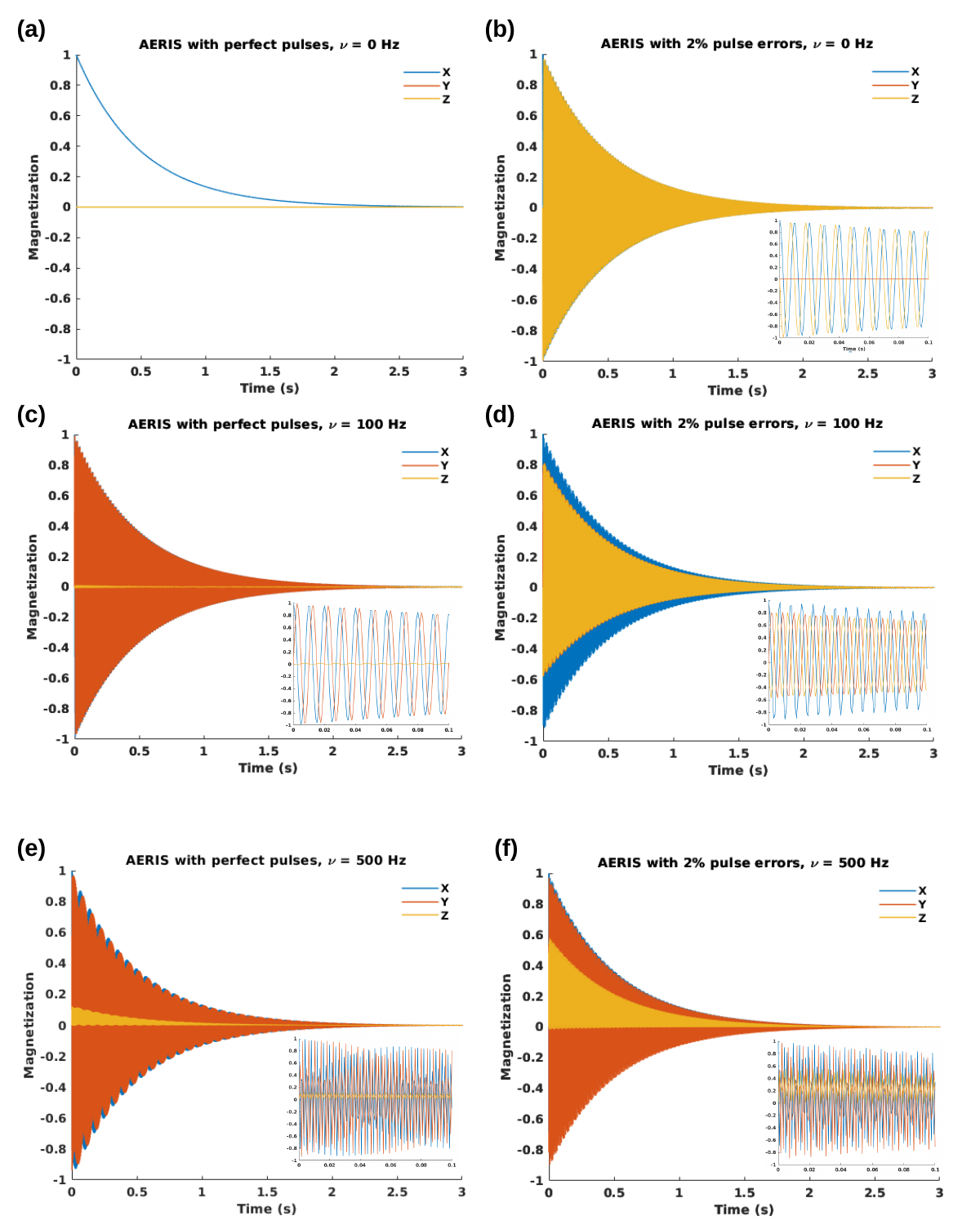}
\captionsetup{singlelinecheck=false,justification=raggedright}
\caption{Time domain simulations of a single nuclear spin during the AERIS sequence help elucidate the dynamics caused by pulse errors. (a) For an on-resonance spin, perfect pulses lead to a simple exponentially decaying signal, the same type of signal that would be acquired with conventional NMR. (b) With 2\% pulse errors, AERIS produces a constant rotation in the X-Z plane that causes an oscillating signal for X-axis magnetization. Spectroscopically, this makes the spin appear to be precessing in the transverse plane at a higher frequency. Since the Y-axis magnetization is not measured with this scheme, it cannot discriminate this effect from actual higher frequency precession in the transverse plane. (c) and (e) When applying perfect pulses to spins detuned by frequency $\nu$, some magnetization oscillates between the longitudinal and transverse planes, but the amount is small, as are the spectral effects. (d) and (f) The effect of pulse errors is smaller for off-resonant spins because the signal is partially refocused as it precesses. For example in (f) the spins spend less time along the Z axis than in (d). For these simulations, $\tau=\mu=800$ ms,  $\Omega = 5000$ Hz, and $T_2 = 1$ s.}
\end{figure*}

\renewcommand{\theequation}{S\arabic{equation}}
\renewcommand\thefigure{S\arabic{figure}}    
\setcounter{figure}{0}  

\end{document}